\documentclass{article}%
\usepackage{graphicx}
\usepackage{amsmath}
\usepackage{amsfonts}
\usepackage{amssymb}%
\begin{document}

\title{Spin Projection Chromatography}
\author{Ernesto P. Danieli, Horacio M. Pastawski* and Patricia R. Levstein\\{\small Facultad de Matem\'{a}tica, Astronom\'{\i}a y F\'{\i}sica,}\\{\small Universidad Nacional de C\'{o}rdoba, }\\{\small Ciudad Universitaria, 5000 C\'{o}rdoba, Argentina.}}
\maketitle

\begin{abstract}
We formulate the many-body spin dynamics at high temperature within the
non-equilibrium Keldysh formalism. For the simplest $XY$ interaction,
analytical expressions in terms of the one particle solutions are obtained for
linear and ring configurations. For small rings of even spin number, the group
velocities of excitations depend on the parity of the total spin projection.
This should enable a dynamical filtering of spin projections with a given
parity i.e. a \textit{spin projection chromatography.} \ 

*corresponding author. \textit{E-mail address}: horacio@famaf.unc.edu.ar (H.
M. Pastawski)


\end{abstract}

\section{Introduction}

The evolution of a \textit{local} spin excitation in a system of interacting
spins\cite{z--Foster,z--deGennes} remains a problem of broad interest in
theoretical and experimental physics. On one extreme, it seeks the
quantification of the hydrodynamic limit of spin \textquotedblleft
diffusion\textquotedblright\ \cite{z--Zeno,z--Cory-spindiff}; on the
other,\ the growing field of quantum information processing needs a
dynamical\cite{z--quant-inf-proc,z--reviewQI-bennett} control of individual
spins in systems of intermediate size pushing the theory to its quantum limit.
In spite of the naive expectation that quantum dynamical interferences might
be limited to simple systems close to its ground state, most of the advances
were originated on Nuclear Magnetic Resonance (NMR)\cite{z--Hahn} which deals
with spin-spin interactions much weaker than the thermal energy $k_{B}T$. The
\textit{discreteness} of the spin system produces \textit{quantum beats }that
allow an efficient cross-polarization among different spin
species\cite{z--MKBE74}. In small \textit{bounded }spin systems, such as
molecules, the initial spreading out of a local excitation is followed by a
revival known as \textit{Mesoscopic Echo}\cite{z--EcosMesoscI} (ME)
manifesting the reflection at the molecular boundaries. This echo was observed
in the proton system of ferrocene\cite{z--EcosMesoscII}, $\mathrm{(C_{5}%
H_{5})}_{2}\mathrm{Fe,}$ and benzene\cite{z--benzene}, $\mathrm{C}%
_{6}\mathrm{H}_{6}$, oriented molecules. Indeed, the complex interferences
produced by the many-body dipolar interactions conspire against the spin wave
behavior and produce a rapid attenuation of the ME. Therefore, a clean
observation of the ME requires a reduction of the complex Heisenberg or
Dipolar Hamiltonians to the simpler \textit{XY } (Planar)
Hamiltonian\cite{z--Madi}. It is now clear, that even when one could control
the many-body dynamics up to a high precision
\cite{z--Exp-Loschmidt,z--Exp-Losch-POWER}, its intrinsic chaotic instability
would easily degrade the quantum coherence\cite{z--JalPa}. This could explain
the fast decay of the Magic Echoes\cite{z--Pines}, the Polarization Echoes
\cite{z--Polar-Echo,z--Polar-Echo-cordoba}, and time reversal of
cross-polarization\cite{z--MatiasErnst} appearing when one \textquotedblleft
inverts\textquotedblright\ the sign of the effective many-spin Hamiltonian.
>From this one learns that an attempt of dynamical control of many spins
systems should focus in the simplest interactions. Spins arranged in a linear
chain with \textit{XY }interaction have a clearly non chaotic dynamics. This
would enable the transfer of the polarization amplitude from one edge to any
specified nucleus in the chain. Besides, as the field of quantum computing
with NMR \cite{z--Paz} grows steadily, one can foresee a new demand for
alternatives to prepare and detect specific quantum states.

The aim of this letter is to propose a novel concept that we call \textit{spin
projection chromatography. }Given a collective spin excitation, it allows to
filter out the components of a given total spin projection parity. To clarify
the physics involved in this phenomenon, we use the known mapping between
spins and fermions \cite{z--XY-dynam-theoI,z--XY-general} to develop a new
formulation for spin dynamics based on the non-equilibrium formalism of
Keldysh\cite{z--Mahan}. Its application to inhomogeneous chains and rings of
spins 1/2 yields closed analytical expressions for the two site spin
correlation functions that facilitate the prediction and understanding of
interference phenomena in the spin dynamics at the molecular level.

\section{Spin Dynamics in the Keldysh Formalism}

The dynamics of a system with $M$ spins under a Hamiltonian $\mathcal{H}$ is
usually described with the two site spin correlation function:%

\begin{equation}
P_{f,i}(t)=\frac{\left\langle \Psi_{eq}\right|  \widehat{S}_{f}^{z}%
(t)\widehat{S}_{i}^{z}(t_{0})\left|  \Psi_{eq}\right\rangle }{\left\langle
\Psi_{eq}\right|  \widehat{S}_{i}^{z}(t_{0})\widehat{S}_{i}^{z}(t_{0})\left|
\Psi_{eq}\right\rangle }. \label{eq--PCFunc}%
\end{equation}
This gives the amount of the $z$ component of the local polarization at time
$t$ on site $f$-th, provided that the system was, at time $t_{0},$ in its
equilibrium state with a spin ``up'' added at $i$-th site. Here, $\widehat
{S}_{f}^{z}(t)=e^{\mathrm{i}\mathcal{H}t}\widehat{S}_{f}^{z}e^{-\mathrm{i}%
\mathcal{H}t}$ is the spin operator in the Heisenberg representation and
$\left|  \Psi_{eq}\right\rangle =\sum_{N}a_{N}\left|  \Psi_{eq}^{(N)}%
\right\rangle $ is the thermodynamical many-body equilibrium state constructed
by adding states with different number $N$ of spins up with appropriate
statistical weights and random phases. The Jordan-Wigner (J-W) transformation
\cite{z--XY-dynam-theoI} establishes the relation between spin 1/2 and
fermion\ operators at site $n$,%

\[
\widehat{S}_{n}^{+}=\widehat{c}_{n}^{+}\exp\{\mathrm{i}\pi\sum_{m=1}%
^{n-1}\widehat{c}_{m}^{+}\widehat{c}_{m}^{{}}\},
\]
with $\widehat{c}_{m}^{+}$ and $\widehat{c}_{m}^{{}},$ the creation and
destruction operator for fermions, and $\widehat{S}_{n}^{\pm}$ the rising and
lowering spin operator $\widehat{S}_{n}^{\pm}=\widehat{S}_{n}^{x}\pm
\mathrm{i}\widehat{S}_{n}^{y}$, where $\widehat{S}_{n}^{\;u}$ ($u=x,y,z$)
represents the Cartesian spin operator. The relation
\[
\widehat{S}_{n}^{z}=\widehat{c}_{n}^{+}\widehat{c}_{n}^{{}}-\tfrac{1}{2},
\]
enables\ us to re-write the numerator of Eq. (\ref{eq--PCFunc}) as a
density-density correlation:
\begin{gather}
\left\langle \Psi_{eq}\right|  \widehat{S}_{f}^{z}(t)\widehat{S}_{i}^{z}%
(t_{0})\left|  \Psi_{eq}\right\rangle =\left\langle \Psi_{eq}\right|
[\widehat{c}_{f}^{+}(t)\widehat{c}_{f}^{{}}(t)\widehat{c}_{i}^{+}%
(t_{0})\widehat{c}_{i}^{{}}(t_{0})\nonumber\\
-\tfrac{1}{2}\widehat{c}_{f}^{+}(t)\widehat{c}_{f}^{{}}(t)-\tfrac{1}%
{2}\,\widehat{c}_{i}^{+}(t_{0})\widehat{c}_{i}^{{}}(t_{0})+\tfrac{1}%
{4}]\left|  \Psi_{eq}\right\rangle . \label{eq--Numerator}%
\end{gather}
Since the last three terms in the rhs are constant occupation factors, the
whole dynamics of the system is given by%

\begin{align}
&  \left\langle \Psi_{eq}\right|  \widehat{c}_{f}^{+}(t)\widehat{c}_{f}^{{}%
}(t)\widehat{c}_{i}^{+}(t_{0})\widehat{c}_{i}(t_{0})\left|  \Psi
_{eq}\right\rangle \\
&  =\sum_{N=0}^{M}\left|  a_{N}^{{}}\right|  ^{2}\sum_{\gamma=1}^{(%
\genfrac{}{}{0pt}{}{M}{N}%
)}\left\langle \Psi_{\gamma}^{(N)}\right|  \dfrac{\mathrm{e}^{-\beta
\mathcal{H}}}{Z_{N}^{{}}}\widehat{c}_{f}^{+}(t)\widehat{c}_{f}^{{}}%
(t)\widehat{c}_{i}^{+}(t_{0})\widehat{c}_{i}(t_{0})\left|  \Psi_{\gamma}%
^{(N)}\right\rangle .\nonumber
\end{align}
Here$,$ $\left|  \Psi_{\gamma}^{(N)}\right\rangle $ and $Z_{N}$ are the
many-body eigenstates and the partition function of the subspace with $N$
particles respectively. We have assumed the high temperature limit,
$k_{B}T=\beta^{-1}\rightarrow\infty$ (i.e. $\mathrm{e}^{-\beta\mathcal{H}%
}\rightarrow I$), that leads to equal statistical weights $\left|
a_{N}\right|  ^{2}=\frac{1}{2^{M}}\left(
\genfrac{}{}{0pt}{}{M}{N}%
\right)  $. Then, the invariance of the trace under cyclic permutations leads to:%

\begin{equation}
\left\langle \Psi_{eq}\right|  \widehat{c}_{f}^{+}(t)\widehat{c}_{f}^{{}%
}(t)\widehat{c}_{i}^{+}(t_{0})\widehat{c}_{i}(t_{0})\left|  \Psi
_{eq}\right\rangle =\left\langle \Psi_{ne}\right|  \widehat{c}_{f}%
^{+}(t)\widehat{c}_{f}^{{}}(t)\left|  \Psi_{ne}\right\rangle .
\label{eq-evol-equil=noequil}%
\end{equation}
The state $\left|  \Psi_{ne}\right\rangle \equiv\widehat{c}_{i}^{+}%
(t_{0})\left|  \Psi_{eq}\right\rangle $ is a \textit{non-equilibrium many-body
state} generated at time $t_{0}$ by creating an excitation at site $i$. The
operator $\widehat{c}_{f}^{+}(t)\widehat{c}_{f}^{{}}(t)$ is evaluated over
this non-equilibrium state. This is a particular case ($t_{1}=t_{2}%
=t$\thinspace and $k=l=f$) of the particle density function\cite{z--Mahan} in
the Keldysh formalism:%

\begin{equation}
G_{k,l}^{<}(t_{2},t_{1})=\tfrac{\mathrm{i}}{\hbar}\left\langle \Psi
_{ne}\right|  \widehat{c}_{k}^{+}(t_{1})\widehat{c}_{l}^{{}}(t_{2})\left|
\Psi_{ne}\right\rangle . \label{eq-density-function}%
\end{equation}
The two-time particle density is related to the exact retarded ($t_{2}>t_{1}$)
\ and advanced ($t_{2}<t_{1}$) Green's functions of the many-body system,
\begin{align}
G_{k,l}^{R}(t_{2},t_{1})  &  =[G_{l,k}^{A}(t_{1},t_{2})]^{\ast}%
\label{eq--GrGa}\\
&  =-\tfrac{\mathrm{i}}{\hbar}\theta\lbrack t_{2}-t_{1}]\left\langle \Psi
_{ne}\right|  \widehat{c}_{k}^{{}}(t_{2})\widehat{c}_{l}^{+}(t_{1}%
)+\widehat{c}_{l}^{+}(t_{1})\widehat{c}_{k}^{{}}(t_{2})\left|  \Psi
_{ne}\right\rangle ,\nonumber
\end{align}
which for quadratic Hamiltonians are just density conserving single particle
propagators. In more complex cases, they can decay through the many-body or
environmental interactions.

Both Eqs. (\ref{eq-density-function}) and (\ref{eq--GrGa}) can be split out
into the contribution of each subspace $N.$ In particular, the non-equilibrium
initial density at $t_{0}$ describing an excitation at site $i$ is,%

\begin{equation}
G_{k,l}^{<_{(N)}}(t_{0},t_{0})=\tfrac{\mathrm{i}}{\hbar}\left(  \tfrac
{N-1}{M-1}\delta_{k,l}+\tfrac{M-N}{M-1}\delta_{k,i}\delta_{i,l}\right)  ,
\label{eq-initial-density}%
\end{equation}
a diagonal matrix involving equal weight for every site and an extra weight at
the excited site $i$. In general, the initial state of Eq.
(\ref{eq-initial-density}) evolves under the Schr\"{o}dinger equation
expressed in the Danielewicz form \cite{z--Danielewicz,z--GLBEII},%

\begin{gather}
G_{f,f}^{\;<_{(N)}}(t,t)=\hbar^{2}\sum_{l,k}G_{f,k}^{R\,_{(N)}}(t,t_{0}%
)G_{k,l}^{\;<_{(N)}}(t_{0},t_{0})G_{l,f}^{A_{(N)}}(t_{0},t)\nonumber\\
+\sum_{l,k}\int_{t_{0}}^{t}\int_{t_{0}}^{t}G_{f,k}^{R_{(N)}}(t,t_{k}%
)\Sigma_{k,l}^{<_{(N)}}(t_{k},t_{l})G_{l,f}^{A_{(N)}}(t_{l},t)\mathrm{d}%
t_{k}\mathrm{d}t_{l}. \label{eq--Danielewicz}%
\end{gather}
The first term in the rhs can be seen as an integral form of the (reduced)
density matrix ($\rho(t)=\mathrm{e}^{-\mathrm{i}\mathcal{H}t/\hbar}%
\rho(0)\mathrm{e}^{\mathrm{i}\mathcal{H}t/\hbar}$) projected over a basis of
single particle excitations in its real space representation as previously
used in related
problems\cite{z--Schmidt-Spiess,z--Fel'dman-cadena,z--Fel'dman-anillo}.
However, the second term would collect incoherent reinjections given by
$\Sigma_{{}}^{\;<}.$ These reinjection can compensate any eventual
\textquotedblleft leak\textquotedblright\ from the coherent evolution whenever
the retarded and advanced propagators, $G_{{}}^{R\,}$ and $G_{{}}^{A\,},$
contain decoherent processes. They also can account for interactions that do
not conserve spin projection. Notice that Eq. (\ref{eq--Danielewicz}) gives
the freedom to operate with two time indices $t_{k}$ and $t_{l},$ containing
information about time correlations \cite{z--GLBEII,z--GLBEI,z--GLBE-luis}.
Hence, one could go beyond the standard initial value solution of the
Schr\"{o}dinger equation and include a continuous coherent injection of
probability amplitude as is required to describe time dependent transport in
mesoscopic and molecular devices\cite{z--GLBE-luis,z--Jauho}. Future
applications may include the descriptions of the chemical kinetics at low
rates of laser pumping into the excited
state\cite{z--Rama-injection,z--Zewail-review,z--Levstein-ChemRev} and the
gradual creation of homonuclear coherence in NMR cross-polarization transfer
\cite{z--MKBE74}.

Equation (\ref{eq--Danielewicz}) allows the evaluation of Eq.
(\ref{eq-evol-equil=noequil}) which gives the non-trivial dynamics of Eq.
(\ref{eq--PCFunc}). Using the high temperature limit, $\left\langle \Psi
_{eq}\right|  \widehat{c}_{f}^{+}(t^{\prime})\widehat{c}_{f}^{{}}(t^{\prime
})\left|  \Psi_{eq}\right\rangle \rightarrow$ $\frac{1}{2},$ the polarization
is written as
\begin{equation}
P_{f,i}(t)=\sum_{N=1}^{M}\dfrac{\left(
\genfrac{}{}{0pt}{1}{M-1}{N-1}%
\right)  }{2^{M-1}}P_{f,i}^{(N)}(t), \label{eq--PCFfin}%
\end{equation}
with%

\begin{equation}
P_{f,i}^{(N)}(t)=\tfrac{2\hbar}{\mathrm{i}}G_{f,f}^{<\,_{(N)}}(t,t)-1.
\label{eq--N_PCFunc}%
\end{equation}
Here, the dynamics $P_{f,i}^{(N)}(t)$ of each subspace $N$, is weighted with
the fraction of states that participate in the evolution.

We are now going to focus into systems where Eq. (\ref{eq--Danielewicz}) takes
the simplest form. If $\mathcal{H}$ commutes with the number operator, as is
the case of the Heisenberg and truncated dipolar Hamiltonians, one is sure
that different subspaces $N$ are decoupled. Further reduction could be
obtained for Hamiltonians which are quadratic in the fermionic operator, such
as the XY in proper conditions, as they might be reduced to non-interacting
particles where $\Sigma_{k,l}^{\;<_{(N)}}\equiv0$. Such cases have fully
coherent evolution and enable the best manifestations of quantum
interferences. Besides, these one-body solutions are needed to build up a
decoherent perturbative approximation of the many-body dynamics.

\section{Solution of Chains and Rings}

Consider a \textbf{linear chain of }$M$\textbf{\ spins}$-\frac{\mathrm{1}%
}{\mathrm{2}}$ in an external magnetic field and interacting with their
nearest neighbors through an $XY$ coupling. Its Hamiltonian,
\begin{equation}
\widehat{\mathcal{H}}^{\mathrm{chain}}=\sum_{n=1}^{M}\hbar\Omega_{n}^{{}%
}\left[  \widehat{S}_{n}^{+}\widehat{S}_{n}^{-}-\tfrac{1}{2}\right]
+\tfrac{1}{2}J_{n+1,n}^{{}}(\widehat{S}_{n+1}^{+}\widehat{S}_{n}%
^{-}+\mathrm{c.c.}). \label{eq--XYchain}%
\end{equation}
has a Zeeman part, $\widehat{\mathcal{H}}_{Z}^{{}}$, proportional to
$\widehat{S}_{n}^{z}$ with $\Omega_{n}^{{}}$ the chemically shifted precession
frequency; and a flip-flop term, $\widehat{\mathcal{H}}_{XY}^{{}},$
where\ $J_{n,n+1}^{{}}$ is the coupling between sites $n$\textit{\ }and
$n+1$\textit{\ }with $J_{M,1}^{{}}=0$.

Due to the short range interaction, after a J-W transformation the only
non-zero coupling terms are proportional to $\widehat{c}_{n+1}^{+}\widehat
{c}_{n}^{{}}=\widehat{S}_{n+1}^{+}\widehat{S}_{n}^{-}$. Each subspace with $({%
\genfrac{}{}{0pt}{}{M}{N}%
})$ states of spin projection $\left\langle \sum_{n=1}^{M}\widehat{S}_{n}%
^{z}\right\rangle =N-M/2$ is now a subspace with $N$ \textit{non-interacting}
fermions. The eigenfunctions $\left\vert \Psi_{\gamma}^{(N)}\right\rangle $
are expressed as a \textit{single} Slater determinant built-up with the single
particle wave functions $\varphi_{\alpha}$ of energy $\varepsilon_{\alpha}$.
Under this condition, $G_{f,i}^{R\,_{(N)}}(t)\equiv G_{f,i}^{R}(t)$ for every
$N,$ and Eq. (\ref{eq--PCFfin}) reduces to
\begin{equation}
P_{f,i}^{\mathrm{chain}}(t)=\hbar^{2}\left\vert G_{f,i}^{R}(t)\right\vert
^{2}. \label{eq--chain-polarization}%
\end{equation}
Even for inhomogeneous systems, this can be solved analytically or numerically
at low computational cost.

\begin{figure}[ptbh]
\centering
\includegraphics[width=7cm]{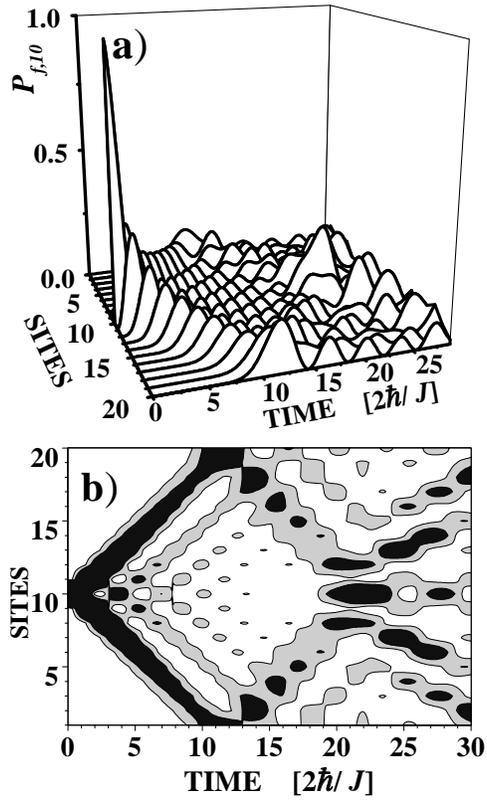}\caption{a) Evolution of the
polarization at each site of a ring of size $M=20$ after a local excitation at
site $10$. b) Contour plot of the polarization. Black filling represents
$P>0.09;$ white is $P<0.045$ and grey corresponds to intermediate values. }%
\label{fig--evo-3D-20spins}%
\end{figure}

For \textbf{spins arranged in a ring}, the closure of the chain requires an
extra term in the Hamiltonian of Eq. \ref{eq--XYchain}, preventing the use of
Eq. (\ref{eq--chain-polarization}). After the J-W transformation this term
takes the form ,%

\begin{equation}
\tfrac{1}{2}J_{M,1}\widehat{S}_{1}^{+}\widehat{S}_{M}^{-}=-\tfrac{1}{2}%
J_{1,M}\widehat{c}_{1}^{+}\widehat{c}_{M}^{{}}\exp[\mathrm{i}\pi\sum_{m=1}%
^{M}\widehat{c}_{m}^{+}\widehat{c}_{m}^{{}}]. \label{eq--WJT1_M}%
\end{equation}
Thus the simple \textquotedblleft flip-flop\textquotedblright\ interaction of
the spin system becomes a \textquotedblleft complex
many-body\textquotedblright\ interaction term in the fermionic representation.
The exponential can take the values $\pm1$ depending on the parity of
$N=\left\langle \Psi_{\gamma}^{(N)}\right\vert \widehat{N}$ $\left\vert
\Psi_{\gamma}^{(N)}\right\rangle ,$ the integer number of particles. Thus,
there are two possible one-body Hamiltonians determining the wave
functions\ $\varphi_{\alpha}(n)$ used to build the Slater determinant
$\Psi_{\gamma}^{(N)}$. All subspaces with the same occupation parity are
described with the same Hamiltonian. The different signs can be seen as a
parity dependent boundary conditions (i.e. either periodic or antiperiodic) or
even better as the effect of an effective potential vector $\mathbf{A}$,
expressed in a singular gauge:%

\begin{equation}
\mathrm{i}\pi\left[  \sum_{m=1}^{M}\widehat{c}_{m}^{+}\widehat{c}_{m}^{{}%
}\right]  _{\operatorname{mod}2}=\mathrm{i}\frac{e}{\hbar}\int_{\mathbf{r}%
_{M}}^{\mathbf{r}_{1}}\mathbf{\mathbf{A}}\mathrm{d}\mathbf{\mathbf{l}%
}=\mathrm{i}\frac{e}{\hbar}%
{\displaystyle\oint}
\mathbf{A}\mathrm{d}\mathbf{l=}\mathrm{i}2\pi\frac{\Phi_{{}}}{\Phi_{0}}.
\label{eq--parity=flux}%
\end{equation}
Here, the second equality defines the singular gauge for $\mathbf{A}$ which
modifies the hopping term according to the Peierls
substitution\cite{z--Feynman}. This results in an Aharonov-Bohm phase
$2\pi\frac{\Phi_{{}}}{\Phi_{0}},$ where $\Phi$ is the \textquotedblleft
fictitious\textquotedblright\ magnetic flux piercing the ring with value
$\Phi=0$ or $\Phi=\frac{1}{2}\Phi_{0}=\frac{1}{2}h/e$ depending on the
particle number (spin projection) parity. Therefore, an \textit{odd
occupation} ($N=2n+1$) yields $\Phi=0$, while the \textit{even occupation}
($N=2n$ with $n$ integer) corresponds to $\Phi=\Phi_{0}/2$. The resulting
Hamiltonian, expressed in the singular gauge and satisfying $\widehat{c}%
_{n}^{{}}=\widehat{c}_{n+M}^{{}}$, reads:%
\begin{equation}
\widehat{\mathcal{H}}^{\mathrm{ring}}(\Phi)=\widehat{\mathcal{H}%
}^{\mathrm{chain}}+\tfrac{1}{2}(J_{1,M}\mathrm{e}^{\mathrm{i}2\pi\Phi/\Phi
_{0}}\widehat{c}_{1}^{+}\widehat{c}_{M}^{{}}+\mathrm{c.c.}).
\end{equation}
The resulting propagators,%
\begin{align}
G_{f,i}^{R\,\,(2n+1)}(t)  &  \equiv G_{f,i}^{R}(t;\Phi=0)~~\mathrm{and}%
\label{eq--G(0)}\\
G_{f,i}^{R\,\,(2n)}(t)  &  \equiv G_{f,i}^{R}(t;\tfrac{1}{2}\Phi_{0}),
\label{eq--G(Pi)}%
\end{align}
are inserted in Eq. (\ref{eq--Danielewicz}) to obtain the polarization from
Eq. (\ref{eq--PCFfin}):
\begin{equation}
P_{f,i}^{\mathrm{ring}}(t)={\tfrac{\hbar^{2}}{2}}\left[  \left\vert
G_{f,i}^{R}(t;\Phi=0)\right\vert ^{2}+\left\vert G_{f,i}^{R}(t;\Phi=\tfrac
{1}{2}\Phi_{0})\right\vert ^{2}\right]  . \label{eq--ring-polarization}%
\end{equation}
Again, the dynamics is completely solved by evaluating the single particle
Green's functions.

\section{Spin Projection and Excitation Velocities}

Let us focus into the spin dynamics of \textit{homogeneous} rings
($J_{n+1,n}\equiv J$ and $\Omega_{n}^{{}}\equiv0$). All the properties can be
expressed in terms of the single particle wave-functions which we solve in the
Coulomb gauge $\mathbf{A}=\frac{1}{2}\mathbf{B}\times\mathbf{r}$ obtaining
$\varphi_{k}(n)=\left\langle 0\right\vert \widehat{c}_{n}^{{}}\widehat{c}%
_{k}^{+}\left\vert 0\right\rangle =\exp[\mathrm{i}\widetilde{k}n]/\sqrt{M}$;
here $\left\vert 0\right\rangle $ is the particle vacuum and $\widehat{c}%
_{k}^{+}$ creates a plane wave with a kinetic momentum $\hbar\widetilde
{k}=\hbar k+\frac{\Phi_{{}}}{\Phi_{0}}\frac{2\pi\hbar}{M}$ obtained from the
stantard discrete wave vector given by $k=s2\pi/M$ with $s=1,\ldots M$; the
lattice constant is $a\equiv1$. $\,$We recall that $\Phi$ is a fictitious flux
identified with the parity of the number of spins up in the original problem.
Hence, a high temperature excitation will require solutions with both fluxes
$\Phi=0$ and $\Phi=\tfrac{1}{2}\Phi_{0}.$

Figure (\ref{fig--evo-3D-20spins}) shows the evolution of a local excitation
through every site of a ring with $M=20$ sites at high temperature. The
original excitation at the $10$-th site $G_{n,m}^{<}(t_{0},t_{0}%
)=\frac{\mathrm{i}}{\hbar}\delta_{n,10}\delta_{10,m}$ splits-out into two wave
packets with opposite velocities (i.e. + or -). Formally, $G_{n,m}%
^{<}(t,t)=G_{n,m}^{<(\mathrm{-})}(t,t)+G_{n,m}^{<(\mathrm{+})}(t,t)\,$where
the \textit{partial} particle densities propagate according to the definite
sign in the kinetic momentum $\hbar\widetilde{k}.$ Let's call $j_{n+1,n}%
(t)=j_{n}^{(\mathrm{+})}(t)-j_{n+1}^{(\mathrm{-})}(t)$ the current density
between sites $n$ and $n+1.$ From Eq. (5.1) in Ref. \cite{z--GLBEII} we
get\ the partial currents:
\begin{equation}
j_{n}^{(\mathrm{\pm})}(t)=\tfrac{1}{2}J_{n,n+1}G_{n\pm1,n}^{<(\mathrm{\pm}%
)}(t,t)-\tfrac{1}{2}J_{n+1,n}G_{n,n\pm1}^{<(\mathrm{\pm})}(t,t),
\end{equation}
defining the propagation in each direction. The components with maximum
velocity become increasingly dominant in the determination of the front of the
wave packet. The eigenenergies $\varepsilon_{\widetilde{k}}=J\cos
[\widetilde{k}]$ allow the evaluation of the group velocity $v_{\widetilde{k}%
}^{{}}=\frac{1}{\hbar}\frac{\partial\varepsilon}{\partial\widetilde{k}}.$ The
partial current and the corresponding local density can be used to evaluate
the mean velocity as $\bar{v}=j/\rho,$ i.e.:
\begin{equation}
\overline{v}(\Phi,t)=\frac{\sum_{n}j_{n}^{(\mathrm{+})}(t)}{\sum_{n}%
\frac{\hbar}{\mathrm{i}}G_{n,n}^{<(\mathrm{+})}(t,t)}=\sum_{0\leq\widetilde
{k}_{2}<\pi/a}\tfrac{2}{M}v_{\widetilde{k}_{2}}^{{}}, \label{eq-mean-velocity}%
\end{equation}
setting a \textit{lower bound} for the peak velocity. The revival time
$T_{ME}$ for the mesoscopic echo is proportional to the ring size $M$ and
inversely proportional to $\overline{v}$ and can be estimated as
$M/v_{\mathrm{\max}}\leq T_{ME}<M/\overline{v}$. In Fig.
(\ref{fig--evo-3D-20spins}b) the dominant group velocity manifests as a linear
dependence of the wave packet maxima with time.

\begin{figure}[ptbh]
\centering \includegraphics[width=\textwidth]{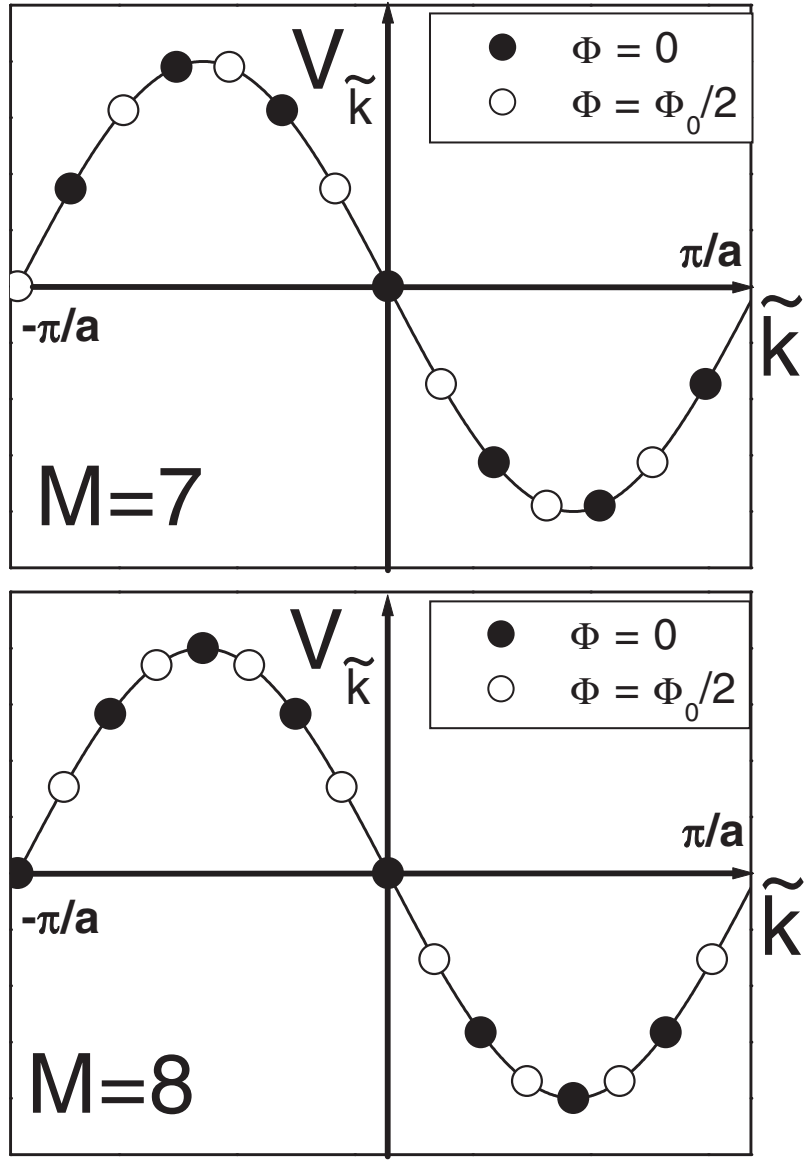}\caption{Dependence of
velocities of eigenstates for rings with \textbf{odd} (7) and \textbf{even
}(8) number of sites as function of the magnetic flux $\Phi$. Full and empty
dots are eigenstates for the cases $\Phi=0$ and $\Phi=\Phi_{0}/2$ which
corresponds to \textbf{odd} \ and \textbf{even} total spin projection parity
respectively. In rings with odd spin number parity the velocities do not
depend on spin projection parity.}%
\label{fig--velocity-7-8spins}%
\end{figure}

There is a fundamental difference between rings of odd and even number of
sites $M.$ For \textit{odd sizes}, the eigenstates determine the \textit{same
set of velocities} regardless of the boundary condition determined by the
Aharonov-Bohm flux (which is half multiple of $\Phi=\Phi_{0}/2$\ or $\Phi=0$).
i.e. neither the available energies nor the velocities depend on the parity of
the particle number, see the top of Fig. (\ref{fig--velocity-7-8spins}).
However, the bottom panel shows that for \textit{even sizes}, this symmetry is
always absent and \textit{the set of velocities depend on} $\Phi,$ i.e. on the
parity of the particle number.

\begin{figure}[ptbh]
\centering \includegraphics[width=10cm]{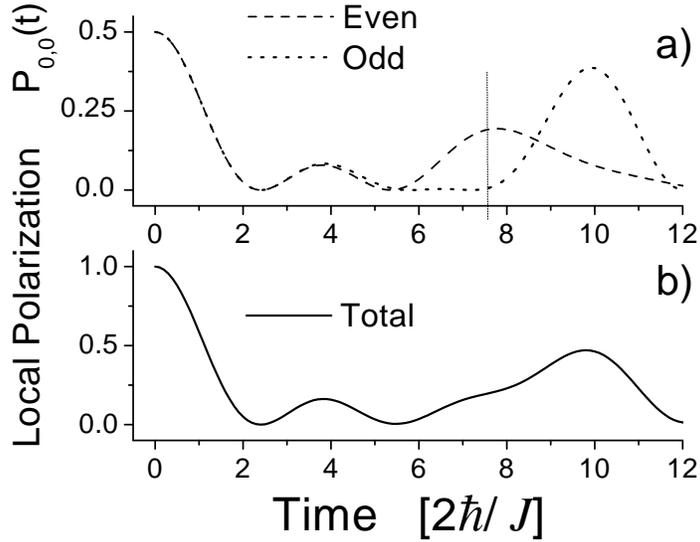}\caption{Spin Projection
Chromatography in a ring with $M=8$. Panel $a)$ shows the contribution of even
and odd spin projections to the local polarization, the sum of both, is shown
in panel $b)$.}%
\label{fig--chromat-8spins}%
\end{figure}

Different velocities for the eigenstates lead to different mean velocities for
the localized excitation. Indeed, the evaluation of Eq.
(\ref{eq-mean-velocity}) for even values of $M$ yields%
\[
\overline{v}(\Phi)=\overline{v}(0)\cos(\frac{\Phi_{{}}}{\Phi_{0}}\frac{2\pi
}{M})-\frac{J}{\hbar}\frac{2}{M}\sin(\frac{\Phi_{{}}}{\Phi_{0}}\frac{2\pi}%
{M}),
\]
while for odd $M$ gives a mean velocity which does not depend on $\Phi$ (i.e.
on the particle parity)
\[
\overline{v}(\Phi)=\overline{v}(0),~~\text{with}~~\overline{v}(0)=-\frac
{J}{\hbar}\frac{2}{M}/\tan(\frac{\pi}{M}).
\]
It is evident that the mean velocity for the even ring sizes satisfies
$\overline{v}(\Phi)>\overline{v}(0)$ indicating a \textit{faster propagation
of the excitations with an even number of particles (i.e. in the subspace with
even spin projection)}. For $M=8,$ $\overline{v}(\frac{1}{2}\Phi
_{0})/\overline{v}(0)=1.08$, in contrast, for $M=7$ this ratio is $1$. This
result is illustrated in Figure (\ref{fig--chromat-8spins}) for an $M=8$ spins
system . Panel $a)$ shows the even ($\widehat{S}_{i}^{z(\mathrm{even})}%
=\sum_{n}\left\langle \Psi^{(2n)}\right\vert \widehat{S}_{i}^{z}\left\vert
\Psi^{(2n)}\right\rangle $) and odd components of the local polarization for
the ring with eight sites. Panel $b)$ shows the total polarization represented
by the sum of both contributions. In summary, \textit{in rings with even
number of spins an excitation in a subspace with even spin projection moves
faster than one in a subspace with odd spin projection}. This conclusion,
easily derived in the language of \textquotedblleft
particles\textquotedblright\ and \textquotedblleft fluxes\textquotedblright,
would have been more elusive in the original language of spins. The novel
concept revealed by the previous simple example is the possibility to
manipulate collective entities as those constituted by the allowed total spin
projections in a many-spin system.

\section{Conclusions}

We must emphasize that the transparency of the formalism fully presented here
for the first time, Eqs. (\ref{eq-density-function}-\ref{eq--PCFfin}), has
already been instrumental to reveal the simple concepts hidden in the various
problems related to spin dynamics such as the nature of Mesoscopic
Echoes\cite{z--EcosMesoscI,z--EcosMesoscII} and Loschmidt Echoes
\cite{z--JalPa}.

The experimental realization of an $XY$ Hamiltonian\ can be achieved through
the truncation of the more complicated Heisenberg (J-coupling) Hamiltonian by
the pulse sequence developed in Ref. \cite{z--Madi}. Its application to
relatively small cyclic molecules\ of even parity (e.g. a derivative of
octatetraene) would enable the observation of the described dynamics. The
evident difference in the delay time of the ME of each spin projection allows
one to manipulate them independently. One could conceive filtering out
components of the total spin projection of a given parity. Consider a local
magnetization (aligned in the $z$ direction) at a site distinguished by its
chemical shift in the $M=8$ ring. This can be achieved by selective
irradiation of one spin or by other polarization transfer method. This
non-equilibrium initial state starts an evolution described by Eq.
(\ref{eq--ring-polarization}). After a time $T_{ME},$ the excitation will
return to the initial site but the components moving in an even spin
projection space will come back first. It is then possible to give a selective
$\pi/2$-pulse (purification pulse) at $t=7.5(2\hbar/J),$ when the even spin
projection has its maximum probability amplitude over the site and the
corresponding probability amplitude for the odd spin projection is still close
to cero, see Fig. (\ref{fig--chromat-8spins}). Thus, the even total spin
projection has been tumbled down towards the $xy-$plane by the purification
pulse leaving the odd one unaltered, i.e. an almost pure odd total spin
projection state lies along the $z-$direction. Hence, the collective nature of
the excitation can be used to achieve a dynamical filtering of the parity of
the total spin allowing the purification of the desired component. This
justifies the name of\textit{\ spin projection chromatography}.

We acknowledge the financial support from ANPCyT, SeCyT-UNC, CONICET and
Fundaci\'{o}n Antorchas.

\end{document}